\documentclass[aps,pra,twocolumn,amsmath,amssymb,preprintnumbers,superscriptaddress,10pt]{revtex4-1}
\usepackage{amsmath}
\usepackage{amsthm}
\usepackage{bm}
\usepackage{graphicx}
\graphicspath{{figures/}}
\usepackage[bookmarks=false]{hyperref}
\usepackage{color}
\usepackage[roman]{sublabel}
\usepackage[capitalise]{cleveref}

\definecolor{pink}{RGB}{255,0,255}

\theoremstyle{definition}
\newtheorem{assumption}{Assumption}

\newcommand{\abs}[1]{\lvert#1\rvert}
\newcommand{\ket}[1]{\left|#1\right>}
\newcommand{\bra}[1]{\left<#1\right|}

\begin{document}

\title{Secure detection in quantum key distribution by real-time calibration of receiver}

\author{{\O}ystein~Mar{\o}y}
\affiliation{Department of Electronic Systems,  NTNU \textendash\ Norwegian University of Science and Technology, NO-7491 Trondheim, Norway}

\author{Vadim~Makarov}
\affiliation{Department of Physics and Astronomy, University of Waterloo, Waterloo, ON, N2L~3G1 Canada}
\affiliation{Institute for Quantum Computing, University of Waterloo, Waterloo, ON, N2L~3G1 Canada}
\affiliation{\mbox{Department of Electrical and Computer Engineering, University of Waterloo, Waterloo, ON, N2L~3G1 Canada}}

\author{Johannes~Skaar}
\affiliation{Department of Electronic Systems,  NTNU \textendash\ Norwegian University of Science and Technology, NO-7491 Trondheim, Norway}
\affiliation{Department of Technology Systems, University of Oslo, Box 70, NO-2027 Kjeller, Norway}

\date{\today}

\begin{abstract}
The single photon detection efficiency of the detector unit is crucial for the security of common quantum key distribution protocols like Bennett-Brassard 1984 (BB84). A low value for the efficiency indicates a possible eavesdropping attack that exploits the photon receiver's imperfections. We present a method for estimating the detection efficiency, and calculate the corresponding secure key generation rate. The estimation is done by testing gated detectors using a randomly activated photon source inside the receiver unit.  This estimate gives a secure rate for any detector with non-unity single photon detection efficiency, both inherit or due to blinding. By adding extra optical components to the receiver, we make sure that the key is extracted from photon states for which our estimate is valid. The result is a quantum key distribution scheme that is secure against any attack that exploits detector imperfections.
\end{abstract}

\maketitle

\section{Introduction}

Quantum key distribution (QKD) \cite{bennett1984, ekert1991} is a method to distribute a secret key between two separate parties, commonly named Alice and Bob. In a QKD scheme Alice and Bob share a quantum channel to distribute the key, as well as an authenticated classical channel for post processing. They also need a source to create a quantum signal, and a detector. An eavesdropper Eve is allowed full control over the quantum channel and may listen to the classical channel. Under these conditions, and under the assumption that Alice and Bob's equipment is flawless, QKD has been proven unconditionally secure \cite{mayers1996, shor2000}.

In the real world equipment is imperfect. Security has been proved for certain general and specific imperfections \cite{gottesman2004, koashi2009, inamori2007}. However, for several different imperfections, attacks against QKD systems have been proposed \cite{makarov2006, lamas-linares2007, zhao2008hacking, xu2010, lydersen2010, lydersen2011superlinear, gerhardt2011full, li2011a, weier2011, jain2011, sun2011, jiang2012, jain2014, sajeed2015a}. Most of these studies experimentally demonstrated imperfection of a system component or subsystem that would allow an attack, but a couple experiments demonstrated successful eavesdropping of the key in a running system \cite{gerhardt2011full, weier2011}. Many realistic attacks, including the latter two, take advantage of imperfections in the detectors \cite{makarov2006, lamas-linares2007, zhao2008hacking, lydersen2010, lydersen2011superlinear, gerhardt2011full, weier2011, jain2011, sajeed2015a, lydersen2011c, tanner2014, kurtsiefer2001, meda2016}. A secure setup requires Bob to measure the signal in a randomly chosen basis. Any differences in detection probability between the bases, in any domain (time, frequency, modes), can be exploited by Eve \cite{fung2009, lydersen2008}. Such differences may either be inherent in the system itself or be forced upon the system by Eve, for example by blinding one of the detectors \cite{lydersen2010, gerhardt2011full}. 

Several solutions have been proposed to the problems caused by imperfect detectors. One option is to use a security proof which is valid for uncharacterized detectors \cite{tomamichel2012tight}; however, then a positive QKD rate requires unrealistically high detection efficiency in the system. A promising approach is the so-called measurement device independent QKD \cite{lo2012}, where a secure key is generated even with untrusted detectors, at the expense of a somewhat more complicated system \cite{rubenok2013, yin2016}. Another, more direct approach, is to find countermeasures for each attack \cite{yuan2010, yuan2011, lim2015}. While convenient and practical, one cannot necessarily be sure that the countermeasures close all types of attacks, or just the already known attacks \cite{lydersen2010reply, lydersen2011comment, huang2016}. Also, the countermeasure itself often requires new components or modified setups, which in turn may open new loopholes.

In this article we will suggest an approach that secures the detector against all attacks as long as some reasonable assumptions are satisfied. This will be done by using the security proof in Ref.~\onlinecite{maroy2010}, where Bob's part of the system is characterized by a parameter $\eta$, which corresponds to the minimum probability that a non-vacuum signal incident to Bob is actually detected by him. Using an additional photon source, we can estimate $\eta$, and quantify the trustworthiness of the detectors. To make sure the assumptions for our proof are satisfied we make some modifications to Bob's part of the system and use bit-mapped gating \cite{lydersen2011}. A secure key rate can then by calculated. Qualitatively, it can be said that the bit-mapped gating and the modifications to Bob takes care of detector efficiency mismatch type loopholes; our additional photon
source takes care of attacks exploiting low single photon detection efficiency, like the blinding attack.

For simplicity we consider the special case of infinite key length (known as the asymptotic limit).  Similarly to the so-called device-independent QKD scenario, we assume that no information leaks out of Alice's and Bob's devices \cite{pironio2009}. However the result may be combined with imperfections in the source, and with information leakage from the detectors, both as done in Ref.~\onlinecite{maroy2010}, and with decoy states \cite{hwang2003, lo2005, wang2005a}.  

\section{Setup and security proof} \label{variable eta}

\begin{figure}
\includegraphics{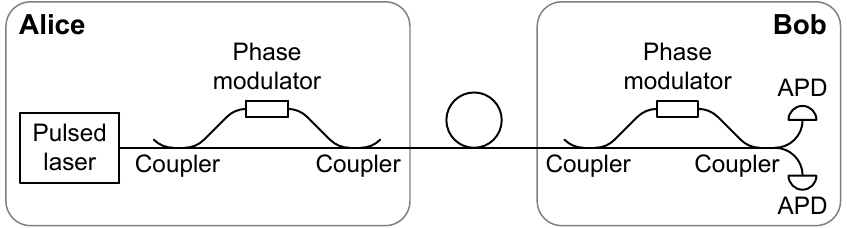}
\caption{\label{Alice&Bob}Alice and Bob's system for BB84 quantum key distribution.}
\end{figure}

We will consider the Bennett--Brassard 1984 (BB84) \cite{bennett1984} protocol using a fiber-based setup with phase coding \cite{townsend1993} and gated detectors. However the ideas presented may also be adapted to free space QKD \cite{buttler1998}, and to other encoding protocols. In phase-based QKD, the key is encoded into the phase difference between the two parts of a light pulse. \Cref{Alice&Bob} shows a typical phase-based QKD system. The pulse is created at Alice's side. Using an unbalanced Mach-Zehnder interferometer she splits it, and encodes her choice of bit and basis by introducing a relative phase shift between the two halves. Bob's part of the system, hereafter referred to as Bob, consists of a similar interferometer and two single photon detectors. 

Alice and Bob choose one of two bases for each pulse and each detection. These two bases are usually referred to as the $z$ basis and $x$ basis. In a practical setup, the pulses live in a large Hilbert space consisting of several Fock spaces, and the $z$ and $x$ bases do not correspond to the usual $z$ and $x$ bases in two-dimensional Hilbert space. They are just names for the two (possibly misaligned) bases which Alice and Bob specify. The key is created from the $n$ pulses where the same basis is chosen by both Alice and Bob, and detection time corresponds to the pulse traveling the short arm of one interferometer and the long arm of the other. For these pulses interference between the two possible paths allows the bit value sent by Alice to be obtained by observing in which of the detectors the pulse arrives. 

The choice of which basis to assign to each letter, $x$ and $z$, is arbitrary. In fact Alice and Bob may randomize this assignment for each pulse, creating a protocol that is symmetric between the bases. The system's average yield $\bar q$, which is the fraction of pulses that are detected by Bob, and $\bar{\delta}$, the average error rate for those pulses, are then both equal in the two bases \footnote{This symmetrization simplifies the analysis of the secret rate. In the case of a difference between yields, $q_z$ and $q_x$, or error rates, $\delta_z$ and $\delta_x$, between the bases of the unsymmetrized protocol, the symmetrization leads to a lower rate. However such differences will generally be small, and the impact on the secret rate insignificant.}. The secure key generation rate extracted from $n\bar q$ bits received by Bob is given by \cite{koashi2009}
\begin{equation} \label{rate}
R=1-h(\bar{\delta})-\frac{H}{n\bar{q}}
\end{equation}
in the asymptotic limit. The function $h(\cdot)$ is the binary Shannon entropy, and $H$ is the amount of privacy amplification needed to remove Eve's knowledge of the key. If $\eta$ is constant during the transmission, we have \cite{maroy2010}
\begin{equation} \label{entropy}
H=n(\bar{q}-\eta \bar{q}(1-h(\bar{\delta}))).
\end{equation}
Here, for simplicity, we have assumed that the source is perfect; the case with imperfections in the source can easily be covered by a small modification to \eqref{entropy} \cite{maroy2010}.

To find a valid numerical expression for the key generation rate, $\eta$ needs to be lower bounded. The parameter $\eta$ is, as in Ref.~\onlinecite{maroy2010}, interpreted as the minimum probability that a non-vacuum incident to the basis dependent interferometer in Bob is detected. Thus $\eta$ is a parameter explicitly given by the state of Bob's system, while the yield $q$ is dependent on both Bob's system and the incoming pulse. In a real QKD experiment the state of Bob's system might change during key exchange. For example, the characteristics of the detectors may change as a result of bright illumination from Eve, as in the blinding attacks \cite{lydersen2010}. We will therefore consider $\eta$ a variable parameter depending on the state of Bob's system. Before each individual qubit measurement  the system is characterized by the parameter $\eta_i$, which is the minimum probability that a non-vacuum signal is detected by Bob. The index $i$ labels the different possible characteristics of Bob's system and $p_i$ is the probability that the system is in the state $i$. Note that $\eta_i$ is independent of Bob's basis choice, it is the minimum over any possible configuration of Bob's system. 

Since Eve may want to tune the yield $q$ and error probability $\delta$, to correlate them with $\eta_i$, we need to index them by $i$ as well. We note that Eve may control $\eta_i$, $q_{i}$, and $\delta_{i}$. In the same way as $\bar\delta$ is the average error probability for those signals that are detected by Bob, we define $\bar{\eta}$ as the average value of $\eta$ for those same detected signals. This is also in accordance with the security analyses in Refs.~\onlinecite{koashi2009,maroy2010}.\footnote{$\bar\eta$ could also have been defined as the average value of $\eta$ over all incoming signals. This would change Eqs. \eqref{Parameter Relations}--\eqref{rate average values} and Eq. \eqref{etaestimate}, but the main result Eq. \eqref{singlephotonkeyrate} would be the same.} According to these definitions the parameters are subject to the relations:
\begin{subequations} \label{Parameter Relations}
\begin{align}
&\sum_i p_i=1,\\
&\sum_i p_iq_{i}=\bar{q},\\
&\sum_i p_iq_{i}\eta_i=\bar{q}\bar{\eta},\label{Eta relation}\\
&\sum_i p_iq_{i}\delta_i=\bar{q}\bar{\delta}.
\end{align}
\end{subequations}
Using random sampling to estimate $\bar{\delta}$, error correction can still be done by sacrificing $h(\bar{\delta})$ bits. The quantity $H$ in \eqref{rate} is now bounded by 
\begin{align} \label{Hboundsum}
H&\leq\max_{p_i,\eta_i,q_{i},q_{i},\delta_{i}}\sum_{i}\left( np_iq_{i}-np_i\eta_i q_{i}(1-h(\delta_{i}))\right) \\\nonumber
&=n\bar q-n\bar\eta\bar q+\max_{p_i,\eta_i,q_{i},\delta_{i}}n\sum_i p_i\eta_i q_{i}h(\delta_{i}).
\end{align}
Using the concavity of the binary entropy, we have, for any $\bar{q}$, $\bar{\eta}$ and $\bar{\delta}\leq {\bar\eta}/{2}$:
\begin{align} \label{inequalety}
\sum_i p_iq_{i}\eta_ih(\delta_{i})&\leq \bar q\bar\eta\sum_i\frac{p_iq_{i}\eta_i}{\bar q\bar\eta}h(\delta_{i})\\\nonumber
&\leq \bar q\bar\eta h\left(\sum_i \frac{p_iq_{i}\delta_{i}\eta_i}{\bar q\bar\eta}\right)\\\nonumber
&\leq \bar q\bar\eta h\left(\sum_i \frac{p_iq_{i}\delta_{i}}{\bar q\bar\eta}\right)\\\nonumber
&\leq \bar q\bar\eta h\left(\frac{\bar\delta}{\bar\eta}\right).
\end{align}
We therefore obtain
\begin{equation} \label{Hmax}
H \leq n\bar q\left(1-\bar\eta\left(1-h\left(\frac{\bar\delta}{\bar{\eta}}\right)\right)\right).
\end{equation}
The bound \eqref{Hmax} is tight as the right hand side is achieved if Eve controls the system as follows: For $np_1q_{1}=n\bar{q}\bar\eta$ detected pulses, the system is in some state with $\eta_1=1$ and $\delta_{1}={\bar\delta}/{\bar\eta}$. For the remaining $n\bar q(1-\bar\eta)$ detected pulses the system is in a state with $\eta_2=0$, and $\delta_{2}=0$.
The key generation rate is bounded by
\begin{equation}\label{rate average values}
 R\geq\bar{\eta}\left(1-h\left(\frac{\bar{\delta}}{\bar{\eta}}\right)\right)-h(\bar{\delta}).
\end{equation}
This rate is similar to the main result from Ref.~\onlinecite{maroy2010}, with the parameter $\eta$ replaced by its average value and $q_x=q_z$. A main difference is the factor $1/\bar\eta$ inside the binary entropy function, which leads to a reduction of the rate. This is due to Eve's ability to avoid introducing errors when the detectors are in a vulnerable state. On the other hand, in Ref.~\onlinecite{maroy2010} $\eta$ must be interpreted as a minimum value, which in many cases leads to zero rate.

\section{Modifications to Bob} \label{modifications}

We now turn to find a lower bound for the average minimum detection probability $\bar\eta$. The main idea is to have a source inside Bob's setup, to test detector sensitivities at random times. We will discuss Bob's parameter $\eta$ as a function of frequency (i.e.,\ wavelength of light), time, or as a function of detailed field distributions. In this context, $\eta$ is defined as the minimum detection probability of a non-vacuum state with the prescribed frequency, time, or field distribution. 

\begin{figure}
\includegraphics{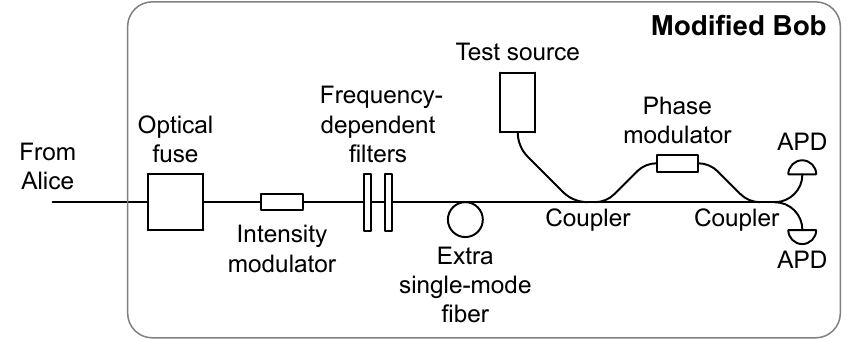}
\caption{\label{ModifiedBob}Bob's modified system including a test source.}
\end{figure}

In general the detection probability depends to some extent on the incoming photons frequency, polarization mode, phase and the time of arrival. These parameters may take infinitely many different values. It is thus unfeasible to find a lower bound for every state experimentally. Instead we restrict the state space $\mathcal{T}$ of the incoming signals by modifying Bob. These modifications will make sure that any single photon pulse in $\mathcal{T}$ will have almost the same detection probability. The modifications are all placed before the interferometer and Bob's basis choice. Any loss due to these components will then only contribute to a smaller yield $q$, and not to $\eta$. We can now estimate $\bar\eta$ from the detection rate $q_T$ of test pulses in $\mathcal{T}$. The test pulses are generated by a source, fired at randomly chosen gates. The pulses are coupled into the fiber scheme before the interferometer. The modifications are shown in \cref{ModifiedBob}. 

The first element in the modified Bob is an optical fuse. If Eve tries to send a pulse with higher power than a certain value $P_0$, this element will be destroyed and communication on the quantum channel will stop. This serves a dual purpose. Bounding the power of Eve's pulse is helpful in our security analysis. Also, deprived of the possibility of sending strong pulses, we may assume that Eve cannot radically change the behavior of the optical elements in Bob's system via laser damage \cite{bugge2014,makarov2016}. 

If some of Eve's pulses are let into Bob when we run the test pulses, our test results will be disturbed. We therefore want an element to deflect, extinguish, or at least dampen Eve's pulses at these times. In combination with the optical fuse, this switch or modulator makes sure that Eve's pulses consist mainly of vacuum when we are sending test pulses. The disturbance of the test measurement statistics must be close to negligible. Note that this switch or modulator should not change the parameters $q_i$ and $\delta_{i}$ in any other pulses from Eve, as this may give Eve some information about when we are sending test pulses.
\begin{assumption} \label{disturbance}
When we are sending a test pulse, any pulse from Eve will change the probability of a detection by at most $\epsilon_E$.
\end{assumption}

To allow just a small bandwidth into Bob, we use a narrowpass filter that transmits light within a frequency range $\Omega=[\omega_0-\omega_B,\omega_0+\omega_B]$, and heavily attenuates all light outside it. In practice, such filter can be achieved by a combination of interference- and absorption-based optical filters. The central frequency of the filter, $\omega_0$, is the same as the central frequency of Alice's pulse \footnote{Another interesting idea is to use frequency dependent phase modulation. Then frequencies different from $\omega_0$ are detected by an increased error rate.}.
\begin{assumption}\label{frequencyzeta}
For any pulse from Eve with frequencies outside the range $\Omega$, the probability that at least one photon is transmitted through the filter and detected is smaller than $q_\omega$ \footnote{Frequencies inside $\Omega$ will also be attenuated by this filter. However, since this loss is basis-independent we can attribute it to Eve and it doesn't contribute to $\eta$.}. 
\end{assumption}
The probability $q_\omega$ depends on the filter performance and the power $P_0$ needed to damage the optical fuse. For later use, let $\zeta_\omega$ be the fraction of detection events that corresponds to photons outside $\Omega$. Clearly, $\zeta_\omega\leq q_\omega/q$. 

For states in $\Omega$ we make the following assumption. 
\begin{assumption}\label{frequency}
For any frequencies $\omega_1,\omega_2 \in\Omega$, $\displaystyle \abs{\eta(\omega_1)-\eta(\omega_2)} \leq \epsilon_\Omega$.
\end{assumption}

When it comes to spatial modes, almost all incoming waves now have the same frequency. We can then insert a short length of single mode fiber in front of the interferometer. 
\begin{assumption}
At most $n\zeta_k$ of the pulses detected by Bob have another mode than the mode allowed in the single mode fiber
\end{assumption} 
The parameter $\zeta_k$ depends on $P_0$ and the length of the single mode fiber. We make sure that all of this fiber is inside Bob so Eve cannot easily access it.

Having made sure that virtually all signals entering Bob have the required frequency and mode, we now consider the timing of the signal. Since Eve controls the fiber between Alice and Bob, we must assume that she can control this timing. A gated detection scheme is thus needed, where the following assumption is satisfied:
\begin{assumption} \label{time} For any times $t_1$, $t_2$ inside the gate, $\displaystyle |\eta(t_1)-\eta(t_2)| \leq \epsilon_T$.
\end{assumption}
As long as this assumption holds true, we can fire our test pulse at any time inside a gate. Eve's advantage by choosing another part of the gate is limited by $\epsilon_T$.

If Bob's system suffers from detector efficiency mismatch \cite{makarov2006}, $\eta$ is not slowly varying at the beginning and the end of the gate. We would therefore like to discard pulses which arrive at these times. Due to jitter in the detector of the same order of magnitude as the detector gate length \cite{cova2004}, it is impossible to recognize these events after detection. We therefore suggest to employ the technique of bit-mapped gating \cite{lydersen2011}. Any signal detected at the beginning and end of the gate will have a random value and contribute to the error rate $\delta$. At least a fraction $(1-2\delta)$ of the detected signals must then have passed inside the inner gate, for which Assumption \ref{time} is valid. Another feature of bit-mapped gating is that the two detectors are randomly assigned to bit value 0 or 1 for each pulse. Thus our setup is equivalent to a setup with one detector which measures whether the bit is 0 or 1.  We therefore don't need to measure $\eta$ for the two detectors seperately. Finally, we note that the detection probability drops to zero immediately after a detection. Although this violates Assumption \ref{time}, it does not affect $\bar{\eta}$ as we can only have one detection per gate.

We also need to consider the number of photons in the pulse. We cannot test all possible states, however for weak pulses detection probability increases with the number of photons in the pulse. Thus the following assumption is natural:
\begin{assumption}\label{photon number}
{The detection probability is smaller for a single photon than for any multiphoton state.}
\end{assumption}

The pulse arrives at Bob in two parts, the first giving detection if the photon travels in the long arm of the interferometer and the last giving detection if the photon travels the short arm. To some extent, $\eta$ may depend on the phase, polarization or detailed shape of the two parts. However, due to assigning each detector to a random bit value for each pulse, such dependency should be small.  Let the detailed field, including the polarization, be described by $\psi(t)$ as a function of time. 

\begin{assumption} For any two distributions $\psi_1(t)$ and $\psi_2(t)$, $\abs{\eta_{\psi_1}-\eta_{\psi_2}} \leq \epsilon_I$. \end{assumption}

\section{Key generation rate with estimation of \bm{$\eta$}}

Because of the modifications to Bob most of the pulses entering the detectors are now in $\mathcal{T}$.  Eve would like as many pulses as possible outside $\mathcal{T}$ since for these pulses she can construct states for which $\eta=0$, without getting noticed. To get an upper bound on the fraction of pulses not in $\mathcal{T}$ we note that being outside of $\mathcal{T}$ in both frequency and mode dimensions further decreases the detection probability of these pulses. Therefore Eve should send pulses which is outside $\mathcal{T}$ only in the dimension where the transmission probability is largest. In addition some of the detected pulses might be outside $\mathcal{T}$ in the time dimension. With probability larger than 
\begin{equation}1-\zeta=1-2\delta-\max\{\zeta_\omega,\zeta_k\}, \end{equation}
a detection event originates from a photon with frequency in $\Omega$, arrival time in the gate, and with the same mode as the test pulse. 

\begin{figure} 
\includegraphics[width=\columnwidth]{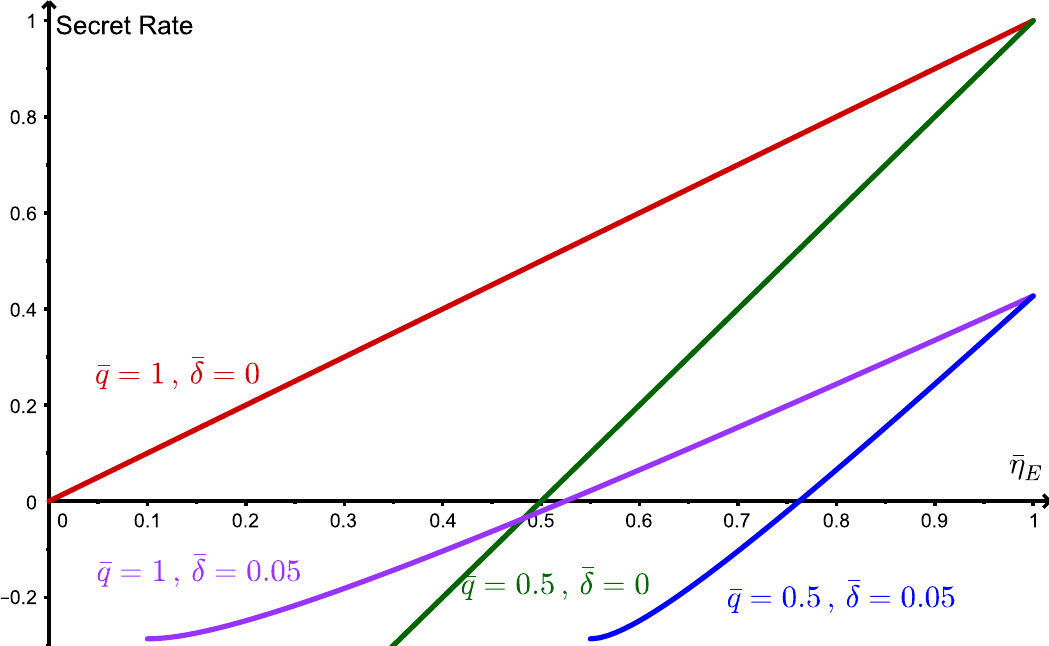}
\caption{\label{fig:Rate}Key generation rate (Eq.~\eqref{singlephotonkeyrate}) as a function of estimated single photon detection efficiency $\bar\eta_E$  for different values of the yield $\bar q$  and error rate $\bar \delta$.}
\end{figure}

If Bob's test pulse is a single photon source, the minimum detection probability ${\eta}_{\mathcal{T}}$ of a single photon state in $\mathcal{T}$ is bounded by the measured detection probability of the test pulse, $q_T$:
\begin{equation} \label{eta q relation}
{\eta}_\mathcal{T}\geq q_T-\epsilon_E-\epsilon_{\Omega}-\epsilon_T-\epsilon_I=q_T-\epsilon_{\text{tot}}.
\end{equation}
Here $\epsilon_{\text{tot}}=\epsilon_E+\epsilon_{\Omega}+\epsilon_T+\epsilon_I$. For states not in $\mathcal{T}$ we have no such bound. A lower bound for the estimated average minimum detection probability of single photon states, $\bar{\eta}_E$, is then
\begin{align}\label{etabound}
 \bar{\eta}_E\geq (1-\zeta)\eta_\mathcal{T} = (1-\zeta)(q_T-\epsilon_\text{tot}).
\end{align}
We now need to take into account that the parameter $\bar\eta$ in Eq. \eqref{rate average values} is the average value of $\eta$ for those $n\bar{q}$ states which were detected. In the worst case scenario, the remaining $n(1-\bar{q})$ non-detected states have $\eta=1$, such that 
\begin{equation} \label{etaestimate}
 \bar{\eta}_E=\bar q\bar{\eta}+(1-\bar q).
\end{equation}
By combining \eqref{rate average values} and \eqref{etaestimate} the expression for the key generation rate when using single photons as test pulses is found to be
\begin{equation} \label{singlephotonkeyrate}
 R\geq\frac{\bar{q}+\bar{\eta}_E-1}{\bar q}\left(1-h\left(\frac{\bar{q}\bar{\delta}}{\bar{q}+\bar{\eta}_E-1}\right)\right)-h(\bar{\delta}).
\end{equation}
This is our main result, in addition to the corresponding expression when using a faint laser source to produce the test pulses (\cref{faint}). The main difference from the rate resulting from \eqref{rate} and \eqref{entropy} is the dependence on the detection rate $\bar q$ and estimated detector parameter $\bar\eta_E$. This is due to the possibility that Eve forces detections when the detectors are in a vulnerable mode and no detections when the detectors are safe. Thus the detectors need to be subject to random testing during key exchange. Any successful attempt from Eve to control the measurement results, without introducing errors, will show up as $\bar\eta_E<1$ during testing. As seen in \cref{fig:Rate}, as long as $\bar q$ remains high, positive key rate is still possible for 
small $\bar\eta_E$. Key gain is possible for $\bar\eta_E+\bar q\geq 1$ in an error-free protocol. 

\section{Key generation rate with practical equipment}
\label{sec:practical-equipment}

When building Bob as described in \cref{modifications} some parts are challenging to implement. In this section we will consider certain solutions to these challenges.

\subsection{Detectors with low efficiency}

\begin{figure} 
\includegraphics[width=\columnwidth]{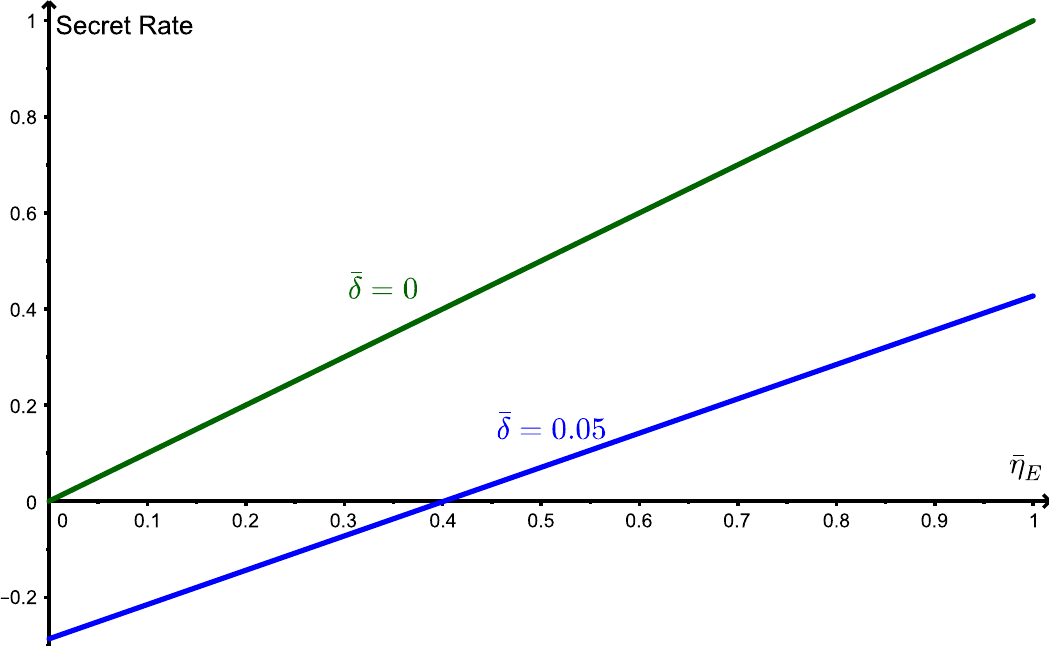}
\caption{\label{fig:Rateetamax}Key generation rate (Eq.~\eqref{rate etamax}) as a function of estimated single photon detection efficiency, $\bar\eta_E$, in the case where $\bar\eta_E$, is equal to the expected value $\eta_\text{max}$. The rate is plotted for different error rates $\bar\delta$, and is valid for any yield $\bar q$.}
\end{figure}

Commercially available detectors used in QKD operate with a detection efficiency substantially less than 1, with most avalanche photodiodes having detection efficiency in the 0.1 to 0.5 range \cite{cova2004,hadfield2009,eisaman2011}. This leads to the factor $(\bar{q}+\bar{\eta}_E-1)$ being negative and \eqref{singlephotonkeyrate} giving negative key rate. A practical solution to this is to assume that no matter what Eve does, she cannot improve the single photon detection efficiency beyond some upper limit $\eta_{\text{max}}$. This leads to some adjustments in the rate. In the third step of Inequality \eqref{inequalety} we now get an extra factor $\eta_\text{max}$ in the argument of the entropy function, 
$\bar q\bar\eta h\left(\sum_i \frac{p_iq_{i}\delta_{i}\eta_i}{\bar q\bar\eta}\right)
\leq \bar q\bar\eta h\left(\sum_i \frac{p_iq_{i}\delta_{i}\eta_\text{max}}{\bar q\bar\eta}\right)$. This factor carries over to Eq. \eqref{rate average values} giving
\begin{equation} \label{rateadj}
 R\geq\bar{\eta}\left(1-h\left(\frac{\bar{\delta}\eta_\text{max}}{\bar{\eta}}\right)\right)-h(\bar{\delta}).
\end{equation}
 Additionally, we must make the following adjustment to Eq. \eqref{etaestimate}: 
\begin{equation}\label{etaestimateadj}
\bar \eta_E=\bar q\bar \eta +(1-\bar q) \eta_\text{max},
\end{equation}
Combining Eqs \eqref {rateadj} and \eqref{etaestimateadj} gives a rate 
\begin{equation}\label{rate etamax}
R\geq\eta_\text{max}\frac{\bar{q}+\frac{\bar{\eta}_E}{\eta_\text{max}}-1}{\bar q}\left(1-h\left(\bar{\delta}\frac{\bar{q}}{\bar{q}+\frac{\bar{\eta}_E}{\eta_\text{max}}-1}\right)\right)- h(\bar{\delta}).
\end{equation}
Note that the expression \eqref{rate etamax} simplifies to
\begin{equation} R=\bar\eta_E(1-h(\bar\delta))-h(\bar\delta) \end{equation}
for $\bar\eta_E=\eta_\text{max}$, i.e. when the detector is working as well as we expect it to. This is the same rate as one gets if $\eta$ is treated as a constant parameter \cite{maroy2010}. This rate is depicted in \cref{fig:Rateetamax} for different error rates $\bar\delta$.  This shows that a detector with low single photon efficiency is not a great security risk in itself, but is detrimental to the rate. A detector which shows a worse single photon efficiency than expected indicates a possible attack from Eve, and requires even more privacy amplification.

\subsection{Estimation of \bm{$\eta$} with a faint pulsed laser}\label{faint} 

\begin{figure} 
\includegraphics[width=\columnwidth]{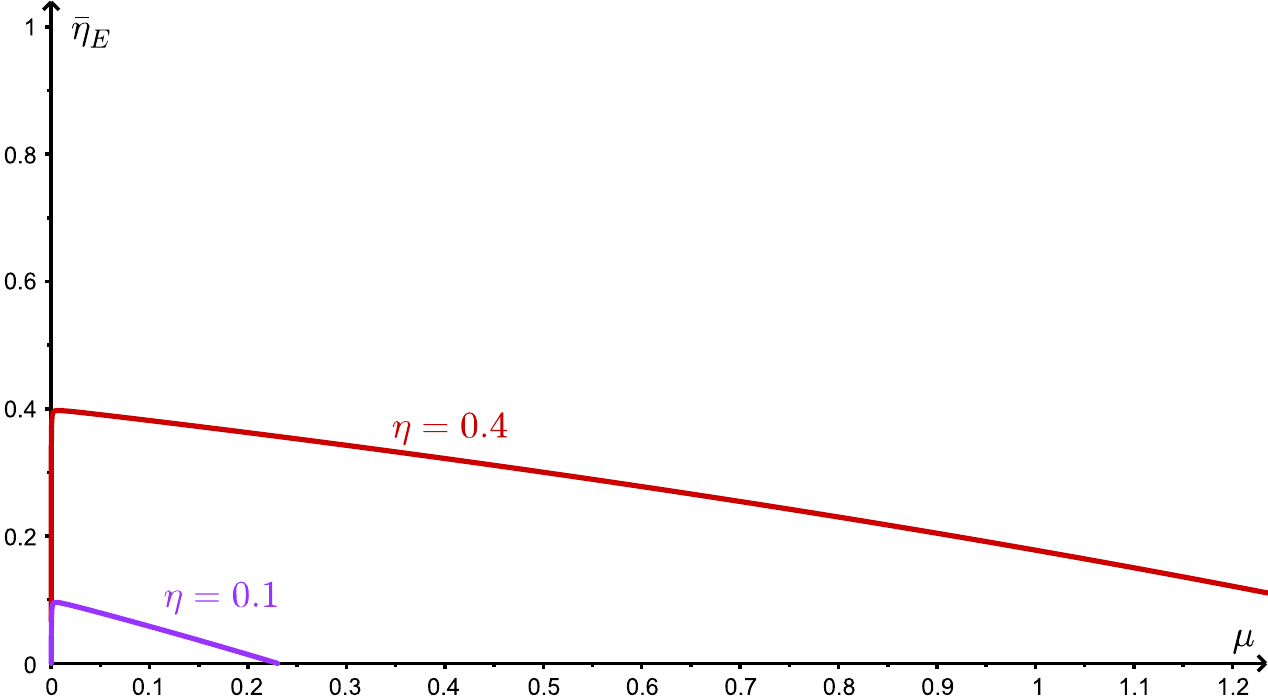}
\caption{\label{fig:faint}Estimated single photon efficiency $\eta_E$ as a function of mean photon number $\mu$ of for two different detectors: One with detection efficiency $\eta=0.1$ and dark count rate $2\cdot 10^{-5}$ (lower line) \cite{stucki2002}, and one with detection efficiency $\eta=0.4$ and dark count rate $2\cdot 10^{-5}$ (upper line).}  
\end{figure}

While single photons sources are available \cite{eisaman2011}, and have been used in some QKD experiments \cite{beveratos2002, alleaume2004}, using a faint laser to produce the test pulse provides an easier setup. For a phase-randomized source with mean photon number $\mu$, the produced state is
\begin{equation}
\rho_B=e^{-\mu}\ket{0}\bra{0}+\mu e^{-\mu}\ket{1}\bra{1}+(1-e^{-\mu}-\mu e^{-\mu})\sigma_m,
\end{equation}
with $\sigma_m$ being all states with more than one photon. This changes the relationship \eqref{eta q relation} between the detection rate of the pulses $q_T$ and the minimum single photon detection probability $\eta_\mathcal{T}$.
We now have
\begin{equation}
 q_T \leq e^{-\mu}d+\mu e^{-\mu}(\eta_\mathcal{T}+\epsilon_\text{tot})+(1-e^{-\mu}-\mu e^{-\mu}),
\end{equation}
or
\begin{equation} \label{eta laser}
 \eta_\mathcal{T} \geq 1-\epsilon_\text{tot}+\frac{1-d}{\mu}-\frac{1-q_T}{\mu e^{-\mu}}.
\end{equation}
In this case the key generation rates given by \eqref{singlephotonkeyrate} and \eqref{rate etamax} are still valid, but the limit for $\bar\eta_E$ in \eqref{etabound} is replaced with
\begin{align}\label{etaEmu}
 \bar{\eta}_E &\geq (1-\zeta)\left(1-\epsilon_\text{tot}+\frac{1-d}{\mu}-\frac{1-q_T}{\mu e^{-\mu}}\right) \nonumber\\
 &= (1-\zeta)\left(\frac{q_T-d}{\mu} + q_T-\frac{\mu}{2}-\epsilon_\text{tot}\right) \nonumber\\
 &~~~\, + \mathcal O(\mu^2) + \mathcal O(\mu q_T).
\end{align}
The dark count rate per gate per pair of detectors, $d$, can be upper bounded by turning off the test pulse while stopping Eve's pulses. As shown in \cref{fig:faint}, for a sufficiently small $d$, we can choose a small $\mu$ and obtain a result which is approximately the same as for a single photon source. The lower line corresponds to a detector with $\eta=0.1,$ $d=2 \cdot10^{-5}$ as in Ref.~\onlinecite{stucki2002}. The upper line shows a more sensitive detector with $\eta=0.4$, $d=2\cdot10^{-5}$.  For both detectors $\zeta=\epsilon_{\text{tot}}=0$ is assumed for simplicity; for multiphoton states each photon is assumed to be detected independently. We see that for both detectors the method described in the previous subsection, using an estimate of $\eta_{\text{max}}$, is needed for positive rate. For the less sensitive detector a very low $\mu$ is needed to approach the true value $\eta=0.1$. 

\subsection{Testing without deflecting Alice/Eve}

The element used to deflect/destroy Alice's pulse during testing must be able to quickly change between total transmittance and total absorbance or deflection. Constructing or finding such an element is challenging. As an alternative, testing can be done without deflecting Alice's pulse, simply by coupling the test pulse into the line using a fiber optic coupler. Assumption \ref{disturbance} is then no longer valid. We can replace it with an assumption bounding the superlinearity \cite{lydersen2011superlinear} of the detector response. 
\setcounter{assumption}{0}
\sublabon{assumption}
\begin{assumption} \label{sublinear detector}
Let $q_T^\prime$ be the actual detection probability under testing. Then $q_T^\prime\leq q+q_T+\epsilon_S$, with $q_T$ being the detection probability if Eve were totally disconnected and $\epsilon_S$ bounding the superlinearity of the detector.
\end{assumption}
The key generation rates given by \eqref{singlephotonkeyrate} and \eqref{rate etamax} are valid in this approach also, with $\epsilon_S$ replacing $\epsilon_E$ and setting $q_T=q_T^\prime-q$ in \eqref{eta q relation}. 

The validity of this assumption needs some further discussion. If a detector is blind for pulses below some threshold intensity and Eve sends pulses slightly below this threshold, clearly $q_T^\prime>q+q_T$ could be possible. However, as long as the test pulse is weak ($\mu<1$), the increase in detection probability by adding a test pulse to Eve's pulse should be small, especially since Eve won't be able to control the exact number of photons in her pulse reaching the detector due to losses in Bob. Therefore $\epsilon_S$ might be considered small.

Security-wise, testing without deflecting Eve is possible, but for the secure key rate such an approach seems disastrous. For positive key rate, $\bar\eta_E+q\geq1$ is needed. The value we use for $\bar\eta_E$ is the lower bound given by \eqref{etabound}. This bound is smaller than $q_T$. For positive rate we therefore need $q+q_T>1$, but this is impossible if we use $q_T=q_T^\prime-q$. Thus either some intensity modulation of Eve's pulse must be done during testing, or stricter assumptions upon the linear response of the detector are needed. 

\section{Discussion and conclusion}

We have proposed a method to estimate the average detection efficiency parameter $\bar \eta$. Given this parameter, QKD is secure for all imperfections in Bob, as long as no signals are emitted from him. Furthermore, information leakage from Alice and Bob can be taken into account by the approach in Ref.~\onlinecite{maroy2010}. In this case, more work is needed to estimate the relevant parameters describing leakage. Alternatively, an approach to estimate leakage is proposed in Refs.~\onlinecite{lucamarini2015,tamaki2016}. The proof of this approach is also based on Koashi's proof \cite {koashi2009} which makes reconciling it with the proof in this article promising. Anyway, the introduction of new components in Bob must be considered in the estimation of information leakage. 

When estimating $\eta$, the parameters in \cref{modifications}, $\zeta_\omega,\ \epsilon_\Omega,\ \zeta_k,\ \epsilon_T$, $\epsilon_I$, and $\epsilon_E$, must be determined by Alice and Bob. If Eve can control these parameters the assumptions of the security proof are not satisfied. The parameters $\zeta_\omega,\ \epsilon_\Omega,\ \zeta_k$, $\epsilon_I$, and $\epsilon_E$ are controlled by the components in Bob and can be estimated by testing these components. Their values are small in a proper setup, and conservative estimates will not affect the key generation rate considerably. We assume that the conservative estimates apply unless irreversible and detectable damage is induced by Eve. Thus these parameters do not need continuous monitoring like $\eta$. To keep $\epsilon_T$ small, a bit-mapped gated detection scheme \cite{lydersen2011} is necessary and sufficient. 

The rate is strongly dependent on the yield $\bar q$ and estimated avreage detection efficiency $\bar\eta_E$. Realistically, the yield is less than the detection efficiency, which means that in practice, $\bar\eta_E\geq 1/2$ is needed for a positive key generation rate.
The reason for this can be seen from the attack where Eve controls the system as described before Equations \eqref{rate average values} and \eqref{etaestimate}. In this case the bound for the secret key generation rate \eqref{singlephotonkeyrate} is tight. Equality is attained by an attack where Eve controls both the incoming pulses and the single photon detection efficiency. In the attack $n\bar q(1-\bar\eta)$ pulses are detected correctly while the detector is blind to single photons. The remaining $n\bar q\bar\eta$ detected pulses are detected, with some errors, while the detectors are sensitive to every single photon pulse. The final $n(1-\bar q)$ pulses are vacuum and the detector would have detected them if they were single photons. This attack shows why a decent detector with single photon efficiency $\eta=0.5$ can be considered insecure. If Eve controls $\eta$ she can e.g. increase it to unity for half of the pulses and let the detector be blind to single photons ($\eta=0$) for the other half. In the instances when the detector is blind she can send weak pulses as in the blinding attack \cite{lydersen2010}, and get full knowledge of the key.  In this case we would still measure $\bar\eta=0.5$. 

To improve the rate, one needs to verify or assume that Eve does not control the system as described before Equations \eqref{rate average values} and \eqref{etaestimate}. We describe one such possible assumption, assuming that Eve cannot increase the single photon detection efficiency beyond some value $\eta_\text{max}$ to fool the estimation procedure. This assumption seems reasonably safe, given that the setup prevents the use of laser damage \cite{bugge2014,makarov2016}. Under this assumption, as long as the estimated single photon $\eta_E$ is close to $\eta_\text{max}$, the key rate before error correction is the same as in QKD with perfect equipment multiplied by the single photon detection efficiency. Key gain is therefore clearly possible, but non-unity single photon detection efficiency is still a disadvantage with respect to the rate.

Implementation of the modified Bob's setup is relatively simple in principle. However, it is challenging to find a sufficiently fast element for deflecting Eve's pulses during testing. Without it, a stricter assumption on the behavior of the detector is needed and the rate suffers. The single photon source in Bob can be replaced by a faint laser similar to the one in Alice, as long as Bob's dark count rate is sufficiently small. 

The setup and method described here can be an important step towards practical secure QKD. Security is no stronger than the weakest link. Since QKD is unconditionally secure with perfect equipment, the implementation is where Eve has had the best opportunities for attack. Our method gives secure key generation under realistic and testable assumptions. Teset

\begin{acknowledgments}
{\O}.M.\ and J.S.\ thank the University Graduate Center, NO-2027 Kjeller, Norway for providing them workplace.
\end{acknowledgments}

%

\end{document}